\documentclass[letter]{aa}  
\usepackage{color}
\usepackage{natbib,twoopt}
\usepackage[hyphenbreaks]{breakurl}
\usepackage[breaklinks]{hyperref}      
\bibpunct{(}{)}{;}{a}{}{,}             
\definecolor{cobalt}{rgb}{0.06, 0.2, 0.65}
\hypersetup{
  colorlinks,
  citecolor=cobalt,
  linkcolor=[rgb]{0.8, 0.2, 1.0},
  urlcolor=cobalt,
}
\makeatletter
  \newcommandtwoopt{\citeads}[3][][]{\href{http://adsabs.harvard.edu/abs/#3}%
    {\def\hyper@linkstart##1##2{}%
     \let\hyper@linkend\@empty\citealp[#1][#2]{#3}}}
  \newcommandtwoopt{\citepads}[3][][]{\href{http://adsabs.harvard.edu/abs/#3}%
    {\def\hyper@linkstart##1##2{}%
     \let\hyper@linkend\@empty\citep[#1][#2]{#3}}}
  \newcommandtwoopt{\citetads}[3][][]{\href{http://adsabs.harvard.edu/abs/#3}%
    {\def\hyper@linkstart##1##2{}%
     \let\hyper@linkend\@empty\citet[#1][#2]{#3}}}
  \newcommandtwoopt{\citeyearads}[3][][]%
    {\href{http://adsabs.harvard.edu/abs/#3}
    {\def\hyper@linkstart##1##2{}%
     \let\hyper@linkend\@empty\citeyear[#1][#2]{#3}}}
\makeatother

\usepackage{graphicx}
\usepackage{txfonts}
\newcommand{\abrame}[1]{}
\newcommand{\rib}[1]{}
\newcommand{\gosic}[1]{}

\begin{document} 
   \title{Complexity-entropy analysis of solar photospheric turbulence: Hinode images of magnetic and Poynting fluxes}
    \titlerunning{Complexity-entropy analysis of solar photospheric turbulence}

   \author{Abraham C.-L. Chian
          \inst{1,2,3}
          \and
          Haroldo V. Ribeiro\inst{4}
          \and
          Erico L. Rempel\inst{1,5}
          \and
          Rodrigo A. Miranda\inst{6}
          \and
          Luis Bellot Rubio\inst{7}
          \and
          Milan~Go\v{s}i\'c\inst{8,9}
          \and
          Breno Raphaldini\inst{10}
          \and
          Yasuhito Narita\inst{1}
          }

   \institute{Institut f\"ur Theoretische Physik, Technische Universit\"at Braunschweig, Mendelssohnstr. 3, 38106 Braunschweig, Germany
                      \and
             School of Computer and Mathematical Sciences, Adelaide University, Adelaide, SA 5005, Australia\\
              \email{abraham.chian@gmail.com}
        \and
            Division of Space Geophysics, National Institute for Space Research (INPE), S\~ao Jos\'e dos Campos, SP 12227‑-010, Brazil
        \and
            Departamento de Física, Universidade Estadual de Maringá, Maringá, PR 87020-900, Brazil
        \and
            Department of Mathematics, Aeronautics Institute of Technology (ITA), S\~ao Jos\'e dos Campos, SP 12228‑900, Brazil
        \and
            Faculty of Science and Technology in Engineering, and Institute of Physics, University of Bras\'ilia, Bras\'ilia, DF 70910‑900, Brazil
        \and
            Instituto de Astrof\'isica de Andaluc\'ia (IAA-CSIC), Apartado de Correos 3004, E-18080 Granada, Spain
        \and
            Lockheed Martin Solar \& Astrophysics Laboratory, 3251 Hanover St, Palo Alto, CA 94304, USA
        \and
            SETI Institute, 339 Bernardo Ave, Suite 200, Mountain View, CA 94043, USA
        \and
            Institute of Mathematics and Statistics (IME), University of S\~ao Paulo, SP 05508-090, Brazil
}

\abstract{The spatiotemporal inhomogeneous-homogeneous transition in the dynamics and structures of solar photospheric turbulence is studied by applying the complexity-entropy analysis to Hinode images of a vortical region of supergranular junctions in the quiet Sun. During a period of supergranular vortex expansion of 37.5 min, the spatiotemporal dynamics of the line-of-sight magnetic field and the horizontal electromagnetic energy flux display the characteristics of inverse turbulent cascade, evidenced by the formation of a large magnetic coherent structure via the merger of two small magnetic elements trapped by a long-duration vortex. Both magnetic and Poynting fluxes exhibit an admixture of chaos and stochasticity in the complexity-entropy plane, involving a temporal transition from low to high complexity and a temporal transition from high to low entropy during the period of vortex expansion, consistent with Hinode observations.}
\maketitle

\section{Introduction}

Photospheric turbulence plays a crucial role in the nonlinear dynamics of solar and stellar plasmas. For example, photospheric vortical flows at the footpoints of magnetic flux tubes/ropes rooted at supergranular junctions can twist magnetic fields, leading to eruption of microflares and flares as well as outflows such as fast solar wind, coronal mass ejections, and switchbacks \citep{Tu2005-lk, Roudier2018-gz, Bale2021-qb}. 

\cite{Gold1960-tk} studied the origin of solar flares by investigating the merger of two twisted magnetic loops of opposite sense and opposite twist. \cite{huang2017magnetospheric} carried out a 2D MHD simulation of two merging magnetic islands to show that the disruption of the current sheet triggers the onset of fast magnetic reconnection. \cite{Rappazzo2019-ia} applied 3D simulation to analyze the twisting of magnetic field lines by photopsheric vortices at the footpoints of two reconnecting magnetic flux tubes and the occurrence of an inverse cascade of magnetic energy. \cite{beg2022Evolution} performed 3D simulation of two merging magnetic flux ropes to show that the initial laminar phase involving the onset of Sweet-Parker reconnection evolves to a transition phase with excitation of secondary plasmoids and then to turbulent magnetic reconnection. \cite{Schiavo2024-qn} applied 2D MHD simulation of magnetic reconnection in two merging flux ropes to show that resistivity plays a central role in regulating the reconnection dynamics. \cite{Sen_2025} used resistive MHD simulations to study the production of small collimated bidirectional nanojet-like ejections due to magnetic reconnection at the interface between two merging plasmoids.

The relation between chaos, stochasticity, complexity, and intermittent turbulence can be studied by computing the permutation entropy \citep{Bandt2002-ti} and the statistical complexity \citep{rosso2007distinguishing}, plotting the results in the complexity–entropy plane. Complexity measures the degree of coherence-incoherence (as well as intermittency and multifractality, see \cite{Miranda2021-xb}); entropy measures the degree of order-disorder. This provides a powerful statistical tool to quantify the degrees of complexity and entropy, which has been applied to characterize the complex nature of space plasma turbulence \citep{Weck2015-zn, Hellinger2019-ts, Olivier2019-gk, Osmane2019-qh, Weygand2019-ub, Good2020-se, Miranda2021-xb, Kilpua2022-ey, Kilpua2024-yw, Raath2022-qv, Bandyopadhyay2024-sf}. These works are based on time series analysis. 

In this paper, we apply the complexity-entropy analysis for the first time to astrophysical images to investigate the incoherent-coherent and disorder-order transitions in the spatiotemporal evolution of photospheric turbulence during a period of vortex expansion, corresponding to a period of build-up of magnetic energy and Poynting flux of the Sun’s relaxations \citep{Chian2024-sx}. In Section 2, we present Hinode observations of the spatiotemporal evolution of photospheric turbulence during a period of vortex expansion, showing patterns of spatiotemporal inhomogeneous-homogeneous transition. In Section 3, we apply the complexity-entropy analysis to study images of photospheric turbulence. In Section 4, we give a discussion and conclusions.

\section{Spatiotemporal evolution of photospheric turbulence during a period of vortex expansion: Inhomogeneous-homogeneous transition}

The nonlinear dynamics of photospheric turbulence has been studied by \hypersetup{citecolor=blue}{\cite{abramenko2001}, \hypersetup{citecolor=blue}{\cite{abramenko2020}}\hypersetup{citecolor=blue}, and \cite{Chian2020-hv, chian2023, Chian2024-sx}. By combining Hinode observation of a photospheric region at a supergranular vertex at the disk center of the quiet-Sun on 2010 November 2-3 and the technique of Lagrangian coherent structures \citep{Rempel2023-os}, these papers have revealed a wealth of information on the complex dynamics and structures of supergranular turbulence, e.g., the objective (i.e., frame invariant) detection of the transport barriers in supergranulations, the center and the boundary of supergranular cells, the interconnection between the centers of adjacent supergranular cells, and long-duration vortices at supergranular junctions. In particular, \cite{chian2023} and \cite{Chian2024-sx} observed the formation of a large coherent magnetic structure resulting from the merger of two small magnetic elements (plasmoids) due to the trapping/untrapping of a supergranular vortex in an interval of vortex expansion-contraction of 60 min; this vortex (see vortex B2 in Table I of \cite{Chian2020-hv}) has a lifetime of 180 min. 

During a period of vortex expansion (associated with a steady increase in vortex area) of 37.5 min, the line-of-sight magnetic field and the horizontal electromagnetic energy flux at the centers of two merging magnetic elements continuously intensify, indicative of the growth of a local small-scale dynamo.  During the period of vortex contraction of 22.5 min, the line-of-sight magnetic field at the center of plasmoid-1 (2) exhibits a steady decrease (increase), respectively, implying a steady transfer of magnetic flux from plasmoid-1 to plasmoid-2. At the end of the vortex expansion–contraction interval, the two merging plasmoids reach an equipartition of electromagnetic energy flux, resulting in the formation of an elongated magnetic coherent structure encircled by a shell of intense current sheets. Evidence of the disruption of a thin current sheet at the interface of two merging plasmoids was observed. In this paper, we will focus on the period of supergranular vortex expansion studied by \cite{Chian2024-sx}.

Figure~\ref{fig1} displays the spatiotemporal evolution of the line-of-sight magnetic field ($B_z$) in the region of a supergranular vertex during the period of vortex expansion studied by \cite{chian2023, Chian2024-sx}, superposed by the Line Integral Convolution (LIC) maps of the horizontal electric current density ($J$) and the objective vortex boundary computed by the instantaneous vorticity deviation \citep{Rempel2023-os}. Since the measurement of the B-vector is not available in our dataset, we obtain the horizontal electric current density from Hinode observations of the line-of-sight magnetic field by assuming vertically oriented magnetic fields, which is a reasonable hypothesis for the footpoints of magnetic flux tubes rooted at supergranular junctions \hypersetup{citecolor=blue}{\citep{Roudier2018-gz, chian2023}}\hypersetup{citecolor=blue}. The cadence of Hinode images is 90 s. The initial timing (Frame 175) of the vortex expansion period is 12:53:51 UT on 2010 November 2; the final timing (Frame 200) of the vortex expansion period is 13:31:21 UT on 2010 November 2. 

\begin{figure}[!t]
    \centering
    \includegraphics[width=0.90\linewidth]{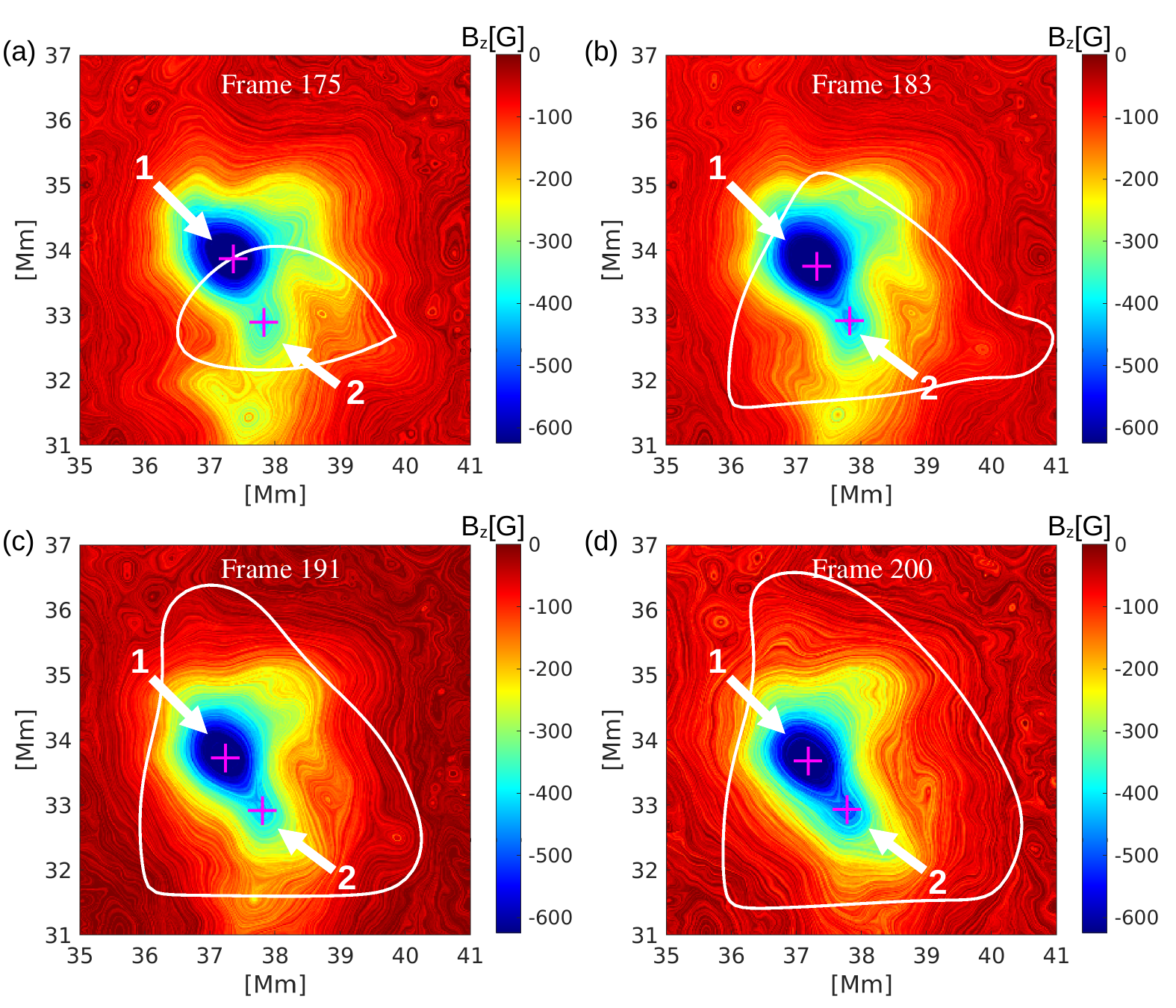}
    \caption{Spatiotemporal inhomogeneous-homogeneous transition of the line-of-sight magnetic field (in $G$) images superposed by the LIC maps of the horizontal electric current density at a supergranular vertex during a period of vortex expansion of 37.5 min. The white line denotes the objective vortex boundary; the arrows denote the merging plasmoids 1 and 2; the magenta crosses denote the centers of plasmoid 1 and 2.}
    \label{fig1}
\end{figure}

Figure~\ref{fig2} displays the spatiotemporal evolution of the horizontal electromagnetic energy flux ($S$) in the same region of Fig.~\ref{fig1} during the period of vortex expansion, superposed by the LIC maps of the horizontal photospheric velocity ($\mathbf{v_h}$) and the objective vortex boundary. The horizontal electromagnetic energy flux $S = (1/4\pi) v_h B_z^2$ is derived from the Poynting vector in the MHD approximation \citep{Silva2022-gr}. 

\begin{figure}[!ht]
    \centering
    \includegraphics[width=0.90\linewidth]{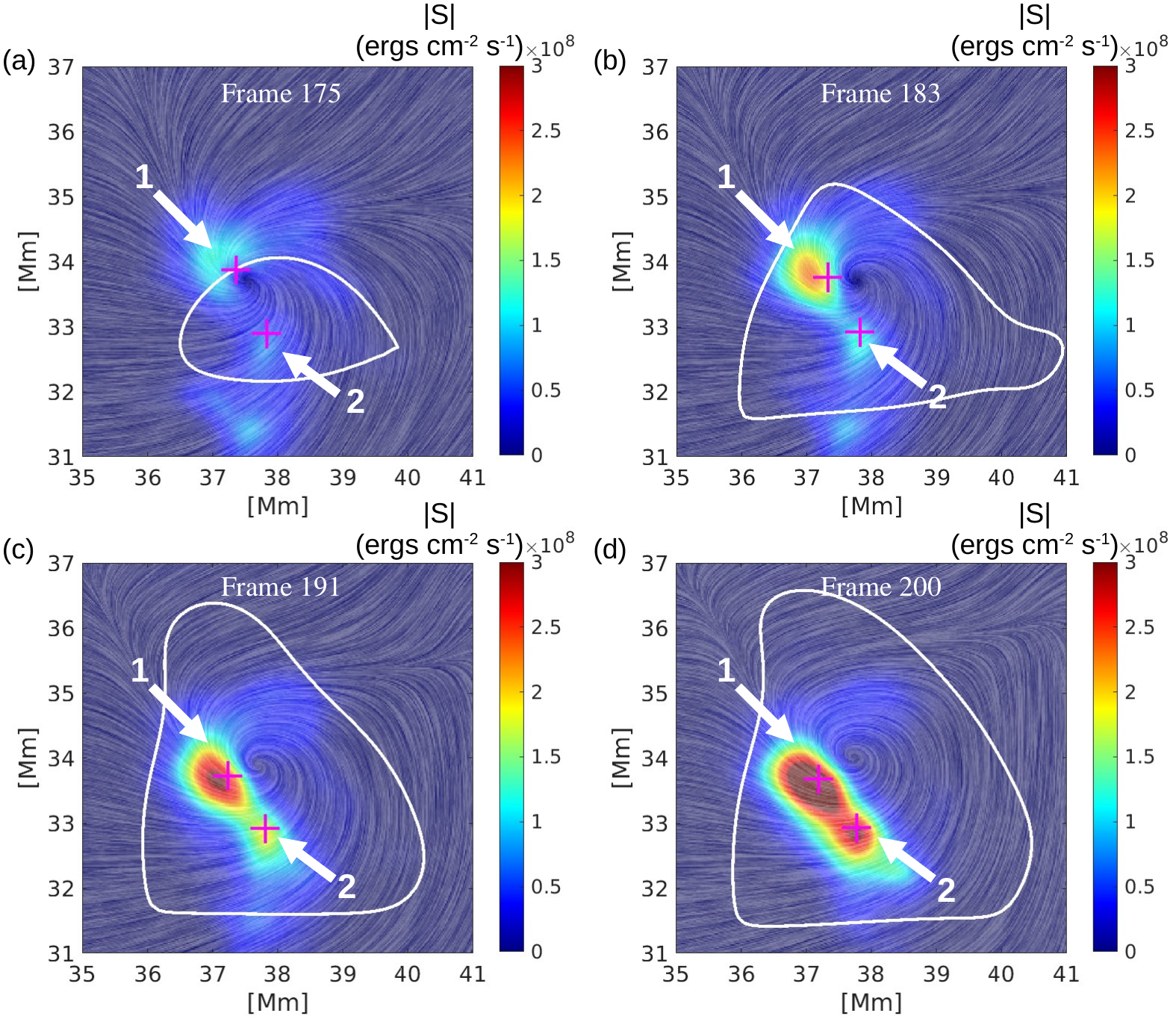}
    \caption{Spatiotemporal inhomogeneous-homogeneous transition of the horizontal electromagnetic energy flux (in ergs cm$^{-2}$ s$^{-1}$) images superposed by the LIC maps of the horizontal photospheric velocity field at a supergranular vertex during a period of vortex expansion of 37.5 min. The white line denotes the objective vortex boundary; the arrows denote the merging plasmoids 1 and 2; the magenta crosses denote the centers of plasmoid 1 and 2.}
    \label{fig2}
\end{figure}

Evidently, Figs.~\ref{fig1} and \ref{fig2} show that there is a steady increase of vortex area during the period of supergranular vortex expansion due to chaotic vortex stretching \citep{chian2023}. At the beginning of the period of vortex expansion (see Frame 175 in Fig.~\ref{fig1}), the distribution of magnetic flux in the region of supergranular vertex is spatially inhomogeneous (i.e., irregular); only a small fraction of the intense magnetic flux concentration is trapped at the interior of vortex boundary. At the end of the period of vortex expansion (see Frame 200 of Fig.~\ref{fig1}), the entire concentration of intense magnetic flux is trapped within the vortex boundary, and an elongated magnetic coherent structure is formed, showing a spatially homogeneous (i.e. regular) distribution of magnetic flux. Note that for the sake of visualization, the images of Figs.~\ref{fig1} and \ref{fig2} are interpolated.

At the beginning of the period of vortex expansion (see Frame 175 of Fig.~\ref{fig2}), the spatial pattern of the horizontal electromagnetic energy flux is inhomogeneous (i.e., irregular), consisting of fragmented patches of Poynting flux; only a fraction of the horizontal electromagnetic energy flux is trapped inside the vortex boundary. At the end of the period of vortex expansion (see Frame 200 of Fig.~\ref{fig2}), the merger of two plasmoids leads to the formation of an elongated electromagnetic coherent structure fully trapped at the interior of the vortex boundary, showing a homogeneous (i.e., regular) spatial distribution of the horizontal electromagnetic energy flux. It follows from Figs.~\ref{fig1} and \ref{fig2} that the spatiotemporal evolution of magnetic and Poynting fluxes demonstrates the complex system characteristics of an inhomogeneous-homogeneous transition resulting from self-organization, transforming from an inhomogeneous state at the beginning of the period of vortex expansion to a homogeneous state at the end of the period of vortex expansion. In the next section, we apply the complexity-entropy approach to investigate this spatiotemporal transition of photospheric turbulence.

\section{Complexity-entropy analysis of images of photospheric turbulence}

The complexity-entropy approaches for analyzing one-dimensional data formulated by \cite{Bandt2002-ti} and \cite{rosso2007distinguishing} were extended by \cite{Ribeiro2012-tn} to two- or higher-dimensional data structures. In this section, we adopt the methodology developed by \cite{Ribeiro2012-tn} to perform a complexity-entropy analysis of Hinode images of photospheric turbulence, using the open-source implementation provided by \cite{Pessa2021-kn}. The analysis is based on the evaluation of ordinal patterns within a two-by-two sliding partition that traverses the image one pixel at a time, both horizontally and vertically. Each ordinal pattern corresponds to one of the twenty-four possible orderings of pixel intensities within the partitions, and their relative frequency defines the ordinal pattern distribution. The normalized Shannon entropy of this distribution yields the normalized permutation entropy ($H$), while the statistical complexity ($C$) is defined in terms of the Jensen-Shannon divergence. Further details of this methodology can be found in \cite{Ribeiro2012-tn} and \cite{Pessa2021-kn}.

We compute the statistical complexity measure and the normalized permutation entropy for Hinode images of the spatiotemporal evolution of the line-of-sight magnetic field and the horizontal electromagnetic energy flux during the period of supergranular vortex expansion given by Figs.~\ref{fig1} and \ref{fig2}. To avoid the introduction of artificial order and complexity from image interpolation, we compute $C$ and $H$ using the raw images of $B_z$ and $S$ of Figs.~\ref{fig1} and \ref{fig2}, respectively, without any interpolation (see Appendix A). Figure~\ref{fig3} shows the computed values of $C$ and $H$ in the complexity-entropy plane for the line-of-sight magnetic field (a) and the horizontal electromagnetic energy flux (b). For the sake of visualization, we show enlarged plots in the inset panels. For comparison, we also include a dashed line representing the values of $C$ and $H$ obtained from two-dimensional fractional Brownian motions (fractal landscapes)~\citep{mandelbrot1982fractal}. It is clear from the enlarged insets that both $B_z$ and $S$ display a temporal transition from low to high complexity and a temporal transition from high to low entropy during the period of vortex expansion. This complexity-entropy characterization of the $B_z$ and $S$ images is consistent with their spatiotemporal evolution seen in Figs.~\ref{fig1} and \ref{fig2}, which show a spatiotemporal inhomogeneous-homogeneous transition during the period of vortex expansion. 

In addition, Fig.~\ref{fig3} shows that magnetic and Poynting fluxes at supergranular junctions exhibit characteristics of an admixture of chaos and stochasticity since the location of the computed values of $C$ and $H$ is above the line of two-dimensional fractional Brownian motion (corresponding to the region 3 in Fig. 1 of \cite{Weygand2019-ub}), albeit indicating higher degrees of complexity and order in comparison with stochastic magnetic fluctuations in the solar wind, whose $C$ and $H$ values lie on top of the line of 1D fractional Brownian motion at the lower right corner of the $C$-$H$ plane (see, e.g., Fig. 7 of \cite{Miranda2021-xb} 2021 and Fig. 7 of \cite{Kilpua2022-ey}). See also Appendix B for other general remarks.

\begin{figure}[!ht]
    \centering
    \includegraphics[width=0.82\linewidth]{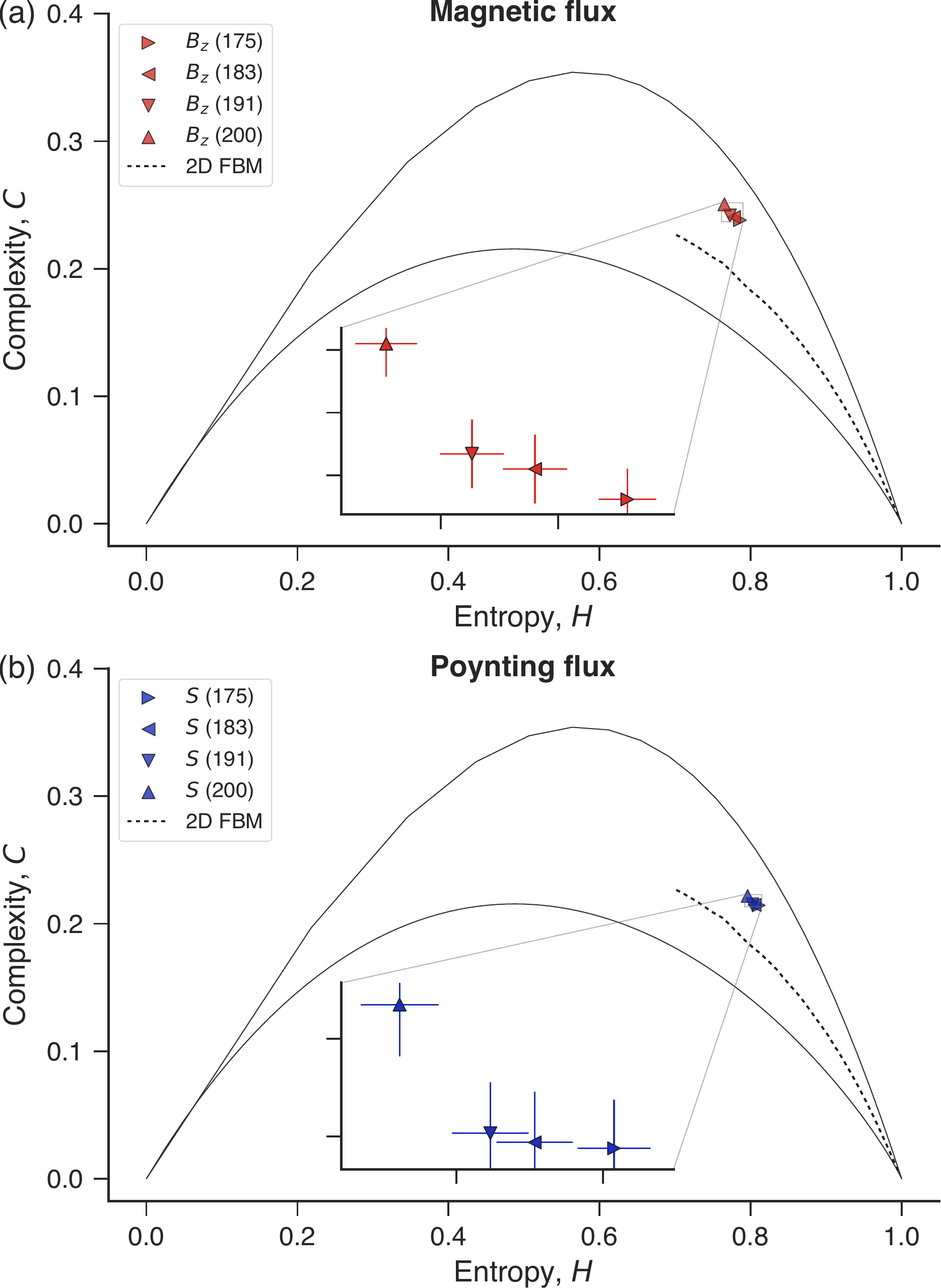}
    \caption{Complexity-entropy plane for the line-of-sight magnetic field $B_z$ (a) and the horizontal electromagnetic energy flux $S$ (b) during a period of supergranular vortex expansion of 37.5 min. The timings of Frames 175, 183, 191, and 200 are indicated by the four respective symbols. The dashed line shows the values of $C$ and $H$ obtained from two-dimensional fractional Brownian motions (2D FBM) with Hurst exponent $h$ ranging from $h\approx0$ (lower right corner) to $h\approx1$; the crescent-shaped curves denote the minimum and maximum values of the complexity, respectively, for a given value of the entropy. Error bars in the insets indicate one-standard-deviation uncertainties of $H$ and $C$ estimated by bootstrapping ordinal patterns and their distributions from the $B_z$ and $S$ images.} 
    \label{fig3}
\end{figure}

\section{Discussion and conclusions}

\cite{Frisch_1975} presented arguments in favor of an inverse cascade of magnetic helicity in MHD turbulence toward small wavenumbers, leading to the generation of magnetic energy at larger scales. \cite{Robinson2023-nf} used the {\it Bifrost} MHD code to investigate the formation and evolution of a pre-flare magnetic flux rope in the quiet Sun and showed that the modeled flux rope is gradually built from a series of component magnetic reconnections involving the coalescence of numerous current-carrying flux tubes; these reconnections lead to an inverse cascade of magnetic helicity from small scales to larger scales. In this paper, we showed in Figs.~\ref{fig1} and \ref{fig2} Hinode observation of the formation of a large magnetic coherent structure (corresponding to the photospheric footpoint of a magnetic flux tube) via the merger of two small plasmoids in conjunction with a local amplification of magnetic field, mediated by a long-duration supergranular vortex, thus supporting the physical mechanisms proposed by \cite{Frisch_1975} and \cite{Robinson2023-nf}.

Most previous complexity-entropy studies of space plasma turbulence \citep{Weck2015-zn, Hellinger2019-ts, Olivier2019-gk, Osmane2019-qh, Weygand2019-ub, Good2020-se, Miranda2021-xb, Kilpua2022-ey, Kilpua2024-yw, Raath2022-qv} found that the fluctuations are stochastic. \cite{Bandyopadhyay2024-sf} reported observational evidence of chaos in the Earth{\textquotesingle s} magnetosheath. In this paper, we adopt the complexity-entropy plane in Fig.~\ref{fig3} to characterize the spatiotemporal inhomogeneous-homogeneous transition of Hinode images of magnetic and Poynting fluxes during a period of supergranular vortex expansion. Our analysis shows that magnetic and Poynting fluxes at supergranular junctions display characteristics of an admixture of chaos and stochasticity. In particular, we elucidated the fundamental nature of spatiotemporal coherent structures by quantifying their high degree of complexity and low degree of entropy. The methodology developed in this paper paves the way for further investigation of complexity-entropy in space and astrophysical images.

\cite{Lapenta2008-gl} conducted MHD simulation to show that magnetic reconnection can progress in two phases: slow laminar Sweet-Parker phase and fast chaotic reconnection phase. The 2D MHD simulation of two merging plasmoids by \cite{huang2017magnetospheric} showed that by varying the Lundquist number on the interface current sheet, reconnection can change from a linear phase to a nonlinear phase of fast reconnection triggered by the current sheet disruption. The 3D simulation of the merger of two magnetic flux ropes by \cite{beg2022Evolution} also showed that reconnection can evolve from a laminar phase to a phase of fast turbulent reconnection. We showed in Fig.~\ref{fig3} that in the initial stage (from Frame 175 to Frame 191, with a duration of 24 min) of the period of supergranular vortex expansion, the increase of the degree of complexity is relatively small and the dynamical changes in this initial stage are slow. This is in agreement with Fig.~\ref{fig1} that shows small dynamical changes of the $B_z$ pattern of two merging plasmoids in this initial stage; moreover, \cite{chian2023} showed that the interface current sheet of two merging plasmoids is weakly perturbed in this initial stage. In contrast, we showed in Fig.~\ref{fig3} that in the final stage (from Frame 191 to Frame 200, with a duration of 13.5 min) of the period of supergranular vortex expansion there is a sharp increase of the degree of complexity and the dynamical changes in this final stage are fast. This is in agreement with Fig.~\ref{fig1} that shows significant dynamical changes of the $B_z$ pattern of two merging plasmoids in this final stage; moreover, \cite{Chian2024-sx} showed that the interface current sheet is strongly perturbed in this final stage. 

To conclude, it follows from the aforementioned discussions that our complexity-entropy analysis was able to confirm the numerical simulations of \cite{Lapenta2008-gl}, \cite{huang2017magnetospheric}, and \cite{beg2022Evolution}, demonstrating that the formation of a magnetic flux tube via the merger of magnetic elements evolves from a slow laminar phase to a fast turbulent phase. Hence, it is likely that the onset of fast turbulent magnetic reconnection may also play a role in the formation of magnetic flux tubes/ropes.

\begin{acknowledgements}
This paper is dedicated to the memory of Professors Vicenzo Carbone, Tom Chang, and Giovanni Lapenta for their pioneering work on intermittent turbulence, complexity, and magnetic reconnection in space and astrophysical plasmas. ACLC expresses gratitude to Professor Narita and Technische Universit\"at Braunschweig for their kind hospitality. The authors thank the Hinode mission for providing the data. HVR and ELR acknowledge support from CNPq (303533/2021-8 and 306920/2020-4). BR and ACLC thank FAPESP for support (23/16513-9 and 24/20546-2). MG was supported by NASA contract NNM07AA01C (Solar-B (Hinode) Focal Plane Package Phase E). RAM acknowledges support from CNPq (407341/2022-6, 407493/2022-0), FAPDF (383/2023) and Brazilian Space Agency (TED 000529/2024).
\end{acknowledgements}

\bibliographystyle{aa_url.bst}
\bibliography{references}
\clearpage

\appendix
\section{The original non-interpolated figures}

Figures~\ref{fig_1appendix} and \ref{fig_2appendix} show the original non-interpolated Hinode images of the line-of-sight magnetic field and the horizontal electromagnetic energy flux used in the complexity-entropy analysis.

\begin{figure}[!ht]
    \centering
    \includegraphics[angle=270, width=\linewidth]{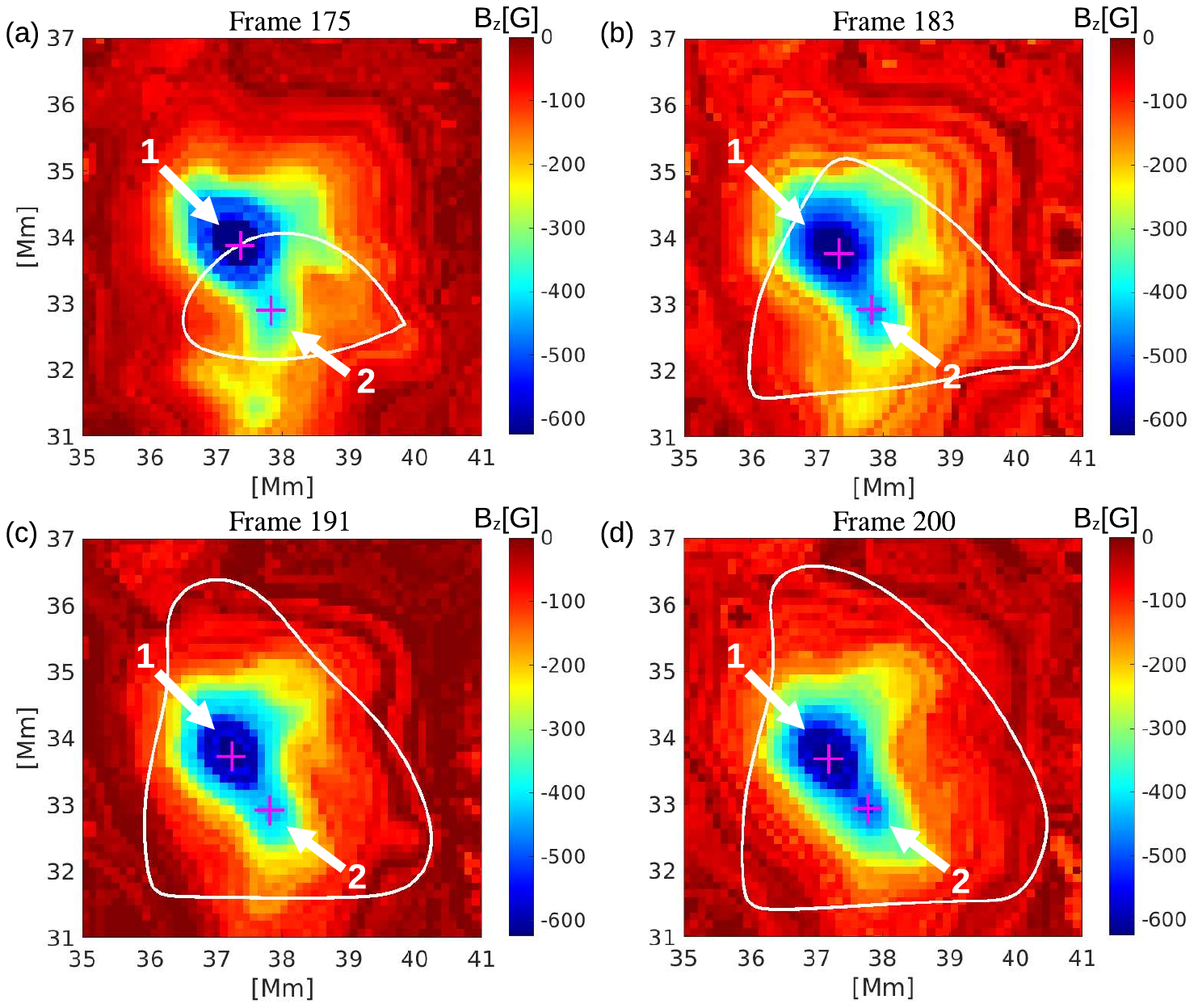}
    \caption{The original non-interpolated Hinode images of the line-of-sight magnetic field of Fig.~\ref{fig1}, with the horizontal electric current density in the background. The white line denotes the objective vortex boundary; the arrows denote the merging plasmoids 1 and 2; the magenta crosses denote the centers of plasmoid 1 and 2.}
    \label{fig_1appendix}
\end{figure}

\begin{figure}[!ht]
    \centering
    \includegraphics[angle=270, width=\linewidth]{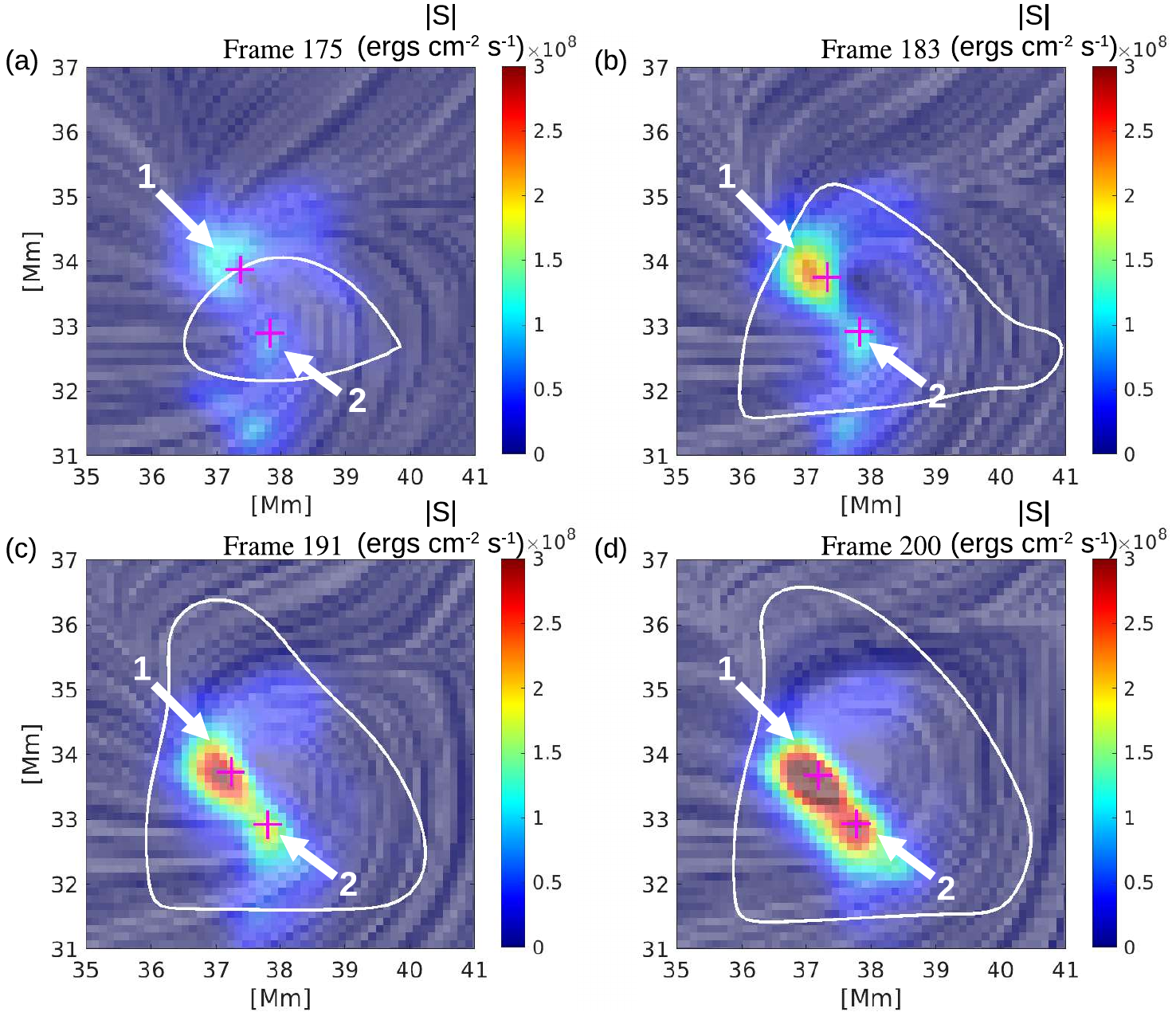}
        \caption{The original non-interpolated images of the horizontal electromagnetic energy flux of Fig.~\ref{fig2}, with the horizontal photospheric velocity field in the background. The white line denotes the objective vortex boundary; the arrows denote the merging plasmoids 1 and 2; the magenta crosses denote the centers of plasmoid 1 and 2.}
    \label{fig_2appendix}
\end{figure}

\section{General remarks}
Magnetic and velocity fields are the fundamental variables of solar photospheric turbulence. It is important to carry out a complex-entropy analysis of both magnetic and velocity fields for the photospheric turbulence. Note, however, that velocity is not an objective (i.e., frame invariant) physical variable. To obtain consistent results, it is necessary to formulate velocity in terms of the Instantaneous Vorticity Deviation, which is objective~\citep{Chian2024-sx}. The results of our ongoing complexity-entropy investigation of the photospheric velocity and magnetic fields will be published in a follow-up paper.

The motivations for analyzing the horizontal electromagnetic energy flux are three-fold. First, \cite{Silva2022-gr} showed through Bifrost numerical simulations and SUNRISE observations that the Poynting flux in the photosphere occurs mainly parallel to the photosphere and contributes to local plasma heating. Second, \cite{chian2023, Chian2024-sx} showed that the horizontal electromagnetic energy flux plays a key role in the formation of magnetic flux tube via plasmoid merger. Third, the horizontal electromagnetic energy flux renders strong support for the complexity-entropy analysis of the line-of-sight magnetic field.

The points in the complexity-entropy serve as a rough guide to distinguishing chaotic or stochastic fluctuations. \cite{ribeiro2017} developed a technique based on the Tsallis $q$-statistics to distinguish accurately whether a fluctuation is chaotic or stochastic, depending on whether the Tsallis complexity-entropy curves are open or closed. We applied this technique to the time interval studied in this Letter and found that these curves are sometimes open and sometimes closed, further supporting our interpretation of the points in Fig.~\ref{fig3} as an admixture of chaos and stochasticity. Note that these points are located in region 3 of Fig. 1 of \cite{Weygand2019-ub}, which they identified as chaotic fluctuations with a strong noisy component. This investigation of the Tsallis complexity-entropy curves will be published in a follow-up paper.

In this Letter, we chose a specific region of supergranular junction studied extensively in a series of papers by \cite{Chian2020-hv, chian2023, Chian2024-sx} to focus on the complex spatiotemporal dynamics of a localized region of strong vortical flows in the photosphere. This particular choice of photosphere provides solid physical insights for interpreting our outcome of the complexity-entropy analysis. The methodology introduced in this Letter may be readily applied to larger domains of the photosphere to improve the validity of statistics.

The resolution of measurement may affect the complexity-entropy evaluations \citep{Kilpua2022-ey, Kilpua2024-yw}. The Hinode data used in our complexity-entropy evaluations are based on a special campaign of the Hinode mission to collect a continuous 22-hour dataset of high spatiotemporal resolution of the quiet-Sun photosphere near the disk center, with high cadence of 90 s and high spatial resolution of 0.32 arcsec per pixel.

In a recent paper by \cite{abramenko2020}, three different regions of the quiet-Sun photosphere were studied using SDO/HMI and the Near InfraRed Imaging Spectrapolarimeter (NIRIS) of Goode Solar Telescope. The HMI magnetic power spectra for the three regions of the quiet-Sun photosphere exhibit the same spectral index of $-1$ computed for a wide range of spatial scales from $10$-$20$ Mm to $2.4$ Mm. A comparison of the HMI and NIRIS spectra with the Kolmogorov-type spectrum showed that more than 35\% of the magnetic energy observed in the scale range of $3.5$–$0.3$ Mm cannot be accounted for by the Kolmogorov turbulent cascade and may be attributed to a local dynamo and/or superdiffusion. The technique of turbulent spectra of \cite{abramenko2020} can be applied to further studies of the photosphere turbulence. It is worth mentioning that the observational evidence of local small-scale dynamo in the quiet-Sun photosphere discussed in this Letter and reported by \cite{chian2023, Chian2024-sx} is consistent with the findings of \cite{abramenko2020}.

Since the measurement of the $B$ vector is not available in the quiet-Sun photosphere, we approximate the $B$ vector by the line-of-sight magnetic field in Fig.~\ref{fig1}. This weakness can be circumvented by reconstructing the $B$ vector using magnetofrictional simulations (see, e.g., \cite{Gosic_2022}).

In this Letter, we demonstrated that the spatiotemporal pattern changes of the inhomogeneous-homogeneous transitions seen in Figs. 1 and 2 can be quantified by the complexity-entropy plane of Fig. 3. In this analysis, entropy measures the degree of order-disorder, ranging from 0 (a perfectly homogeneous image in which a single ordinal pattern occurs) to 1 (a completely noisy image in which all ordinal patterns are equally likely). In turn, complexity quantifies the strength of correlational structure (``structuredness'') in the system and approaches zero at both extremes of order and disorder, reflecting the intuitive idea that high complexity lies between these limits. Moreover, complexity is not a trivial function of entropy: for a given entropy value, complexity can vary between a minimum and a maximum (as indicated by the continuous curves in Fig. 3), thus providing information not captured by entropy alone. This provides a powerful statistical tool for quantifying the degrees of complexity and entropy in the inhomogeneous-homogeneous transitions of astrophysical turbulence.

The enlarged insets in Fig.~\ref{fig3} for both $B_z$ and $S$ display a monotonic temporal transition from low to high complexity and a monotonic temporal transition from high to low entropy. This is in agreement with the monotonic temporal evolutions of the objective vortex area, the total Instantaneous Vorticity Deviation in vortex, and the maximum value of the Instantaneous Vorticity Deviation during the time interval of vortex expansion (see Fig. 3 of \cite{Chian2024-sx}), as well as in agreement with the monotonic temporal evolutions of the total line-of-sight magnetic field at the interior of vortex boundary, and of the modulus of the line-of-sight magnetic field and the horizontal electromagnetic energy flux at the centers of two merging plasmoids (see Fig. 7 of \cite{Chian2024-sx}). The consistency of these results is an indication of the robustness of our complexity-entropy analysis.

\end{document}